\documentclass[aps,prb,twocolumn,floatfix]{revtex4}
\usepackage{times}
\usepackage{graphicx}
\usepackage{epsfig}
\usepackage{amsmath}
\usepackage{amssymb}
\usepackage[version=3]{mhchem}
\usepackage{nicefrac,xfrac}

\begin{document}

\title{Efficient fitting of single-crystal diffuse scattering in interaction space: a mean-field approach}

\author{Ella M. Schmidt}
\affiliation{Department of Chemistry, University of Oxford, Inorganic Chemistry Laboratory, South Parks Road, Oxford OX1 3QR, U.K.}

\author{Johnathan M. Bulled}
\affiliation{Department of Chemistry, University of Oxford, Inorganic Chemistry Laboratory, South Parks Road, Oxford OX1 3QR, U.K.}

\author{Andrew L. Goodwin$^\ast$}
\affiliation{Department of Chemistry, University of Oxford, Inorganic Chemistry Laboratory, South Parks Road, Oxford OX1 3QR, U.K.}

\date{\today}
\begin{abstract}
The diffraction patterns of crystalline materials with strongly-correlated disorder are characterised by the presence of structured diffuse scattering. Conventional analysis approaches generally seek to interpret this scattering either atomistically or in terms of pairwise (Warren--Cowley) correlation parameters. Here we demonstrate how a mean-field methodology allows efficient fitting of diffuse scattering directly in terms of a microscopic interaction model. In this way the approach gives as its output the underlying physics responsible for correlated disorder. Moreover, the use of a very small number of parameters during fitting renders the approach surprisingly robust to data incompleteness, a particular advantage when seeking to interpret single-crystal diffuse scattering measured in complex sample environments. We use as the basis of our proof-of-concept study a toy model based on strongly-correlated disorder in diammine mercury(II) halides.
\end{abstract}


\maketitle
	\section{Introduction}
	
Complex structures can emerge from simple interactions \cite{Ziman_1979,Parsonage_1978,Welberry_1985}. Geometric frustration in the Ising triangular antiferromagnet \cite{Wannier_1950}, the hydrogen-bonding-driven configurational degeneracy of cubic ice \cite{Bernal_1933}, and the long-period stacking phases of the anisotropic next-nearest-neighbour interaction (ANNNI) model \cite{Bak_1982} are all well-studied examples. It is a natural corollary that complexity is not particularly uncommon, and indeed there is a growing realisation that complexity of various types is not only present but important for the behaviour of many key classes of functional materials --- from disordered rocksalt cathodes to high-temperature superconductors \cite{Clement_2020,Ji_2019,Mydosh_2011,Welberry_2016,Simonov_2020b}. Determining the structures of such systems is one of the key challenges of modern structural science \cite{Billinge_2007,Keen_2015,Juhas_2015}.

\begin{figure}
\includegraphics[width=0.95\columnwidth]{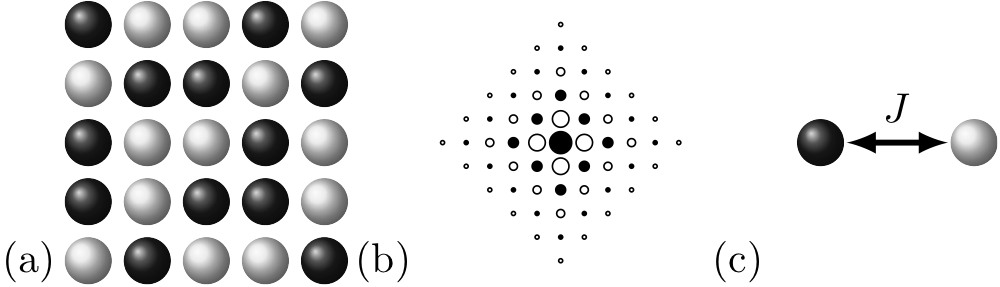}
\caption{Descriptions of complex structures, illustrated for the case of a disordered binary alloy. (a) Representative atomistic configuration. (b) Auto-correlation function, with positive maxima represented by filled circles, and negative maxima represented by empty circles. The size of the circle corresponds to the strength of the auto-correlation. (c) Interaction space driving the generation of atomistic models as in (a) and the corresponding auto-correlation function in (b).}
\label{fig1}
\end{figure}

Implicit in the term `complex' is the inference that a very large number of parameters is needed to describe a structure meaningfully. In the case of disordered crystals, for example, one approach is to use atomistic configurations, each representing a structural fragment spanning sufficiently many unit-cells to capture any key correlations; the corresponding number of descriptors is large indeed because it scales with the supercell volume and is amplified further by the loss of crystal symmetry \cite{McGreevy_1988,Eremenko_2019,Goodwin_2019} [Fig.~\ref{fig1}(a)]. An alternative is to describe disordered structures in terms of interatomic correlations --- such as the Warren--Cowley parameters --- since these are uniquely determined for disordered crystals even if they can be realised by many microscopically distinguishable configurations \cite{Cowley_1950,Weber_2012}. Yet even these parameters are many because correlations usually extend across a large number of unit cells, there are many individual contributions to the correlation function for a given pair, and one needs in principle to consider not only two-body, but also higher-order terms [Fig.~\ref{fig1}(b)]. Hence there is an apparent paradox in that some complex structures can be succinctly generated, but not succinctly described.

An obvious resolution is to describe complex structures in terms of the (simple) interactions from which they arise --- we will come to call this an `interaction space' description. Doing so is the motivation, of course, of direct Monte Carlo (MC) studies of disordered and other complex materials: one tests candidate interaction models by comparing their predictions against observables, until the best model --- the `solution' --- is identified \cite{Weber_2005,Welberry_1985} [Fig.~\ref{fig1}(c)]. In a crystallographic context, the key experimental observable one might use to best discriminate different models is the structured diffuse scattering patterns measured in suitable single-crystal diffraction experiments \cite{Neder_2008,Welberry_2014}. This diffuse scattering can be calculated directly from MC configurations, allowing a quantitative measure of goodness-of-fit. A combination of (i) varying the interaction parameter(s), (ii) re-running the corresponding MC simulations, and (iii) assessing the change in quality of fit-to-data then forms the basis of an interaction-space refinement strategy, such as employed in the inverse Monte Carlo (IMC) \cite{Weber_2005,Almarza_2003,Jain_2006b,DAlessandro_2011} and empirical potential structure refinement (EPSR) \cite{Soper_1996,Soper_2012} approaches.

Here we explore the viability of a particularly efficient alternative method for refining interaction parameters against diffuse scattering data that bypasses altogether the need to generate atomistic configurations \emph{en route} [Fig.~\ref{fig1}(c)]. Not only is the approach computationally attractive, but it removes the uncertainty introduced by employing a stochastic method such as MC. The approach itself is based on mean-field theory, and has been applied previously in various guises to the study of plastic crystals and frustrated magnets \cite{Naya_1974,Nagai_1982,Descamps_1982,Derollez_1990,Enjalran_2004,Paddison_2013}. We anticipate that a suitably generalised methodology could be of enormous value in the investigation of disordered crystals beyond these two specific cases.

Our paper is arranged as follows. We begin by presenting the underlying theory for the mean-field calculation of single-crystal diffuse scattering from a given interaction model. In section~\ref{procrystal} we briefly introduce the disordered physical system that will form the basis of our proof-of-principle study, and develop a related, simplified, two-dimensional toy model. In both cases we present the corresponding single-crystal diffuse scattering patterns and set as our challenge the task of recovering from this scattering the underlying physics responsible for driving complexity. Section~\ref{results} sets out the results of our mean-field analysis, which we compare against the results of conventional approaches based on reverse Monte Carlo (RMC) and Warren--Cowley (WC) methodologies. We consider in turn both the 2D toy model and its 3D (physical) parent. This section concludes with a discussion of the relative sensitivities of different approaches to data loss; a key result is that the mean-field approach is remarkably robust in this respect. In the final section~\ref{conclusions} we summarise the opportunities and challenges for generalising this mean-field methodology to enable the systematic investigation of complexity in a wide range of different materials.

\section{Theory}\label{theory}
    
    Mean-field theory is a self-consistent field theory, widely used in statistical physics to model high-dimensional random systems \cite{Curie_1895,Weiss_1907,Kadanoff_2009}. Here, we present the formalism as applied to orientationally disordered molecular crystals, but will come to show how this interpretation might equally well be applied to compositional disorder.

    \subsection{Mean-field formalism}
    
Our starting point is to define a suitable pair-interaction Hamiltonian for our particular system of interest. We consider the system as comprised of individual building blocks (\emph{e.g.}\ molecules), which (i) can adopt any one of a fixed number of discrete orientations, (ii) are positioned on a periodic lattice, and (iii) interact with one another in a pairwise sense. The generalised Hamiltonian is then (following Ref.~\citenum{Naya_1974}): 
    \begin{equation}
        \mathcal{H} = \frac{1}{2} \sum_{j} \sum_{k} \sum_{l=1}^{s} \sum_{m=1}^{s} \mu_{j}^{l} J^{lm}_{jk}\mu_{k}^{m},
        \label{eq:Hamilton1}
    \end{equation}
    where $j$ and $k$ sum over all unit cells in the crystal, and $l$ and $m$ sum over all $s$ possible orientations of the disordered molecule. 
    The variables $\mu_{j}^{l}$ are equal to $1$ if the molecule at site $j$ is in orientation $l$, and $0$ otherwise; the $J^{lm}_{jk}$ are the components of the pair-interaction Hamiltonian.
    
    It will be convenient for us to express Equation \eqref{eq:Hamilton1} in matrix form:
    \begin{equation}
        \mathcal{H} = \frac{1}{2} \sum_{j} \sum_{k} \boldsymbol{\mu}_{j} \underline{\underline{J}}_{jk}\boldsymbol{\mu}_{k},
        \label{eq:Hamilton2}
    \end{equation}
    where $\boldsymbol{\mu}_{j}$ is the $s$-dimensional vector with components $\mu_{j}^{l}$ and $\underline{\underline{J}}_{jk}$ is an $s\times s$-dimensional interaction matrix.
    
    Working within the same formalism, the diffuse scattering intensity $I(\boldsymbol{q})$ is given by
    \begin{equation}
        I(\boldsymbol{q}) \propto \mathrm{Tr} \left\{ \underline{\underline{F}} \left< \boldsymbol{\mu}_{\boldsymbol q} \cdot\boldsymbol{\mu}_{\boldsymbol q}    \right> \right\}.
        \label{eq:ID1}
    \end{equation}
    Here $\underline{\underline{F}}$ is the $s\times s$-dimensional matrix of molecular form factors and $\boldsymbol{\mu}_{\boldsymbol q}$ is the Fourier transform of $\boldsymbol{\mu}_{j}$. The brackets $\left<\right>$ denote the expectation value.
    
    The mean field approximation as taken by Ref.~\citenum{Naya_1974} allows us to express the diffuse scattering intensity in terms of the pair-interaction Hamiltonian:
    \begin{equation}
    		I(\boldsymbol{q}) \propto\mathrm{Tr}\left\{\underline{\underline{M}}\underline{\underline{F}}\left[\underline{\underline{1}} + \beta \underline{\underline{M}}\underline{\underline{J}}(\boldsymbol{q}) \right]^{-1}\right\},
		    \end{equation}
   where $\underline{\underline{M}}$ is the $s\times s$ matrix of average orientational populations, with elements
   \[ M_{ij} = \frac{1}{s}\delta_{ij} - \frac{1}{s^{2}} \]
   for the case that all $s$ orientations occur with probability $1/s$.
   $\underline{\underline{J}}(\boldsymbol{q})$  is the Fourier transform of the pair interaction matrix, and $\beta = \frac{1}{k_{\rm B}T}$ is the inverse thermodynamic temperature.
   
   This expression can be recast as an eigenvalue problem
   \begin{equation}
   I(\boldsymbol{q})=\gamma\sum_{i=1}^{s} - \frac{\left[\underline{\underline{U}} \underline{\underline{M}} \underline{\underline{F}} \underline{\underline{U}}^{-1} \right]_{ii}}{1+\beta \lambda_{i}},
    		\label{eq:ID2}
		\end{equation}
where $\underline{\underline{U}}$ is the matrix that transforms $\underline{\underline{M}}\underline{\underline{J}}(\boldsymbol{q})$ into diagonal form and $\lambda_{i}$ are the eigenvalues of $\underline{\underline{M}}\underline{\underline{J}}(\boldsymbol{q})$. The parameter $\gamma$ is a scale factor. Note that the contribution of a given eigenmode to the scattering function is numerically greatest when the corresponding eigenvalue is large and negative.

By assigning a cost function $\chi^2=\sum_{\boldsymbol q}[I_{\rm exp}({\boldsymbol q})-I({\boldsymbol q})]^2$ to quantify the difference between experimental and mean-field diffuse scattering intensities, one can use conventional least-squares approaches to refine the entries in the interaction matrices $\underline{\underline{J}}_{jk}$.

   \subsection{Approximations in the mean field derivation}
   
   The approach by Ref.~\citenum{Naya_1974} is developed in the so-called random phase approximation.
   In this approximation, the $\mu_{j}^{l}$ that describe the occupation of the disordered sites in the crystalline system are treated as random variables.
   The expectation value $\left< \mu_{j}^{l}\right>$ is given by the average occupation of the molecular component as determined by average structure analysis.
   For the systems presented here, this average occupation is constrained by the average symmetry and is constant for all different orientations.
   
   The local correlations and the constraint that in the real, disordered system, each site is occupied by exactly one orientation, collectively imply that the $\mu_{j}^{l}$ are not independent random variables.
   In the mean-field approximation, the central limit theorem is applied and it is assumed that the probability distributions of $\boldsymbol{\mu}_{\boldsymbol{q}}$ is given as an $s$-dimensional Gaussian probability distribution \cite{Naya_1974}:
   \begin{equation}
     P(  \boldsymbol{\mu}_{\boldsymbol{q}}) =  \frac{\exp\left(-\frac{1}{2} \boldsymbol{\mu}_{\boldsymbol{q}}^{\rm T} \underline{\underline{M}}_{\boldsymbol{q}}^{-1} \boldsymbol{\mu}_{\boldsymbol{q}} \right)}{\sqrt{(2\pi)^{s}\det \left(\underline{\underline{M}}_{\boldsymbol{q}}  \right)}},
   \end{equation}
   with $\underline{\underline{M}}_{\boldsymbol{q}}$ the matrix of the second moments of the random variable $\boldsymbol{\mu}_{\boldsymbol{q}}$.
   
   The random phase approximation --- and hence our mean-field analysis --- is expected to break down for ordered phases, such as an antiferromagnetic Ising system at low temperatures.
   This is the case when the pair-interaction energies are large with respect to the available thermal energy.
   Ref.~\citenum{Naya_1974} gives the criterion
   \begin{equation}
         \det \left[ \underline{\underline{1}} + \beta \underline{\underline{M}}\underline{\underline{J}}(\boldsymbol{q}) \right] \ge 0
        \label{MFstability}
    \end{equation}
    to describe the temperature range in which the mean-field approximation might be expected to give physical results.  In the related mean-field formalism developed for frustrated magnets \cite{Enjalran_2004}, an equivalent criterion is used that depends on the most negative eigenvalue $\lambda_{\rm{min}}(\boldsymbol{q})$ in Equation \eqref{eq:ID2}:
     \begin{equation}
         \beta \le -\lambda_{\rm {min}}(\boldsymbol{q}).
        \label{MFstability2}
    \end{equation}
    Hence, for all refined pair-interaction energies in the mean field approximation presented here, it has to be checked that both inequalities \eqref{MFstability} and \eqref{MFstability2} are satisfied.
    
    \subsection{Number of included pair-interaction terms}
    
In principle, there is no \emph{a priori} limit on the number of pair-interaction terms that might be included in the interaction matrix \eqref{eq:Hamilton2} --- although our motivation of attempting to describe complex structures succinctly implies we wish in due course to use as few terms as possible. If too many parameters are used, and their values subsequently refined against experimental diffuse scattering data as outlined above, then such refinements are often unstable. This instability can often be traced back to violation of the criteria in \eqref{MFstability} and/or \eqref{MFstability2}.

A suitable strategy, in our view, is as developed by Ref.~\citenum{Paddison_2013} for the identification and refinement of magnetic interactions in the frustrated magnet $\beta$-Co$_x$Mn$_{1-x}$. Pair-interaction terms are included in a refinement procedure one-by-one and the goodness of fit monitored. Only terms that lead to a significant improvement can be interpreted in terms of a physical interaction model and should be included in the subsequent analysis.        
    
\section{Model system}\label{procrystal}    

\subsection{Parent compound}

\begin{figure*}
        \includegraphics[width=0.85\textwidth]{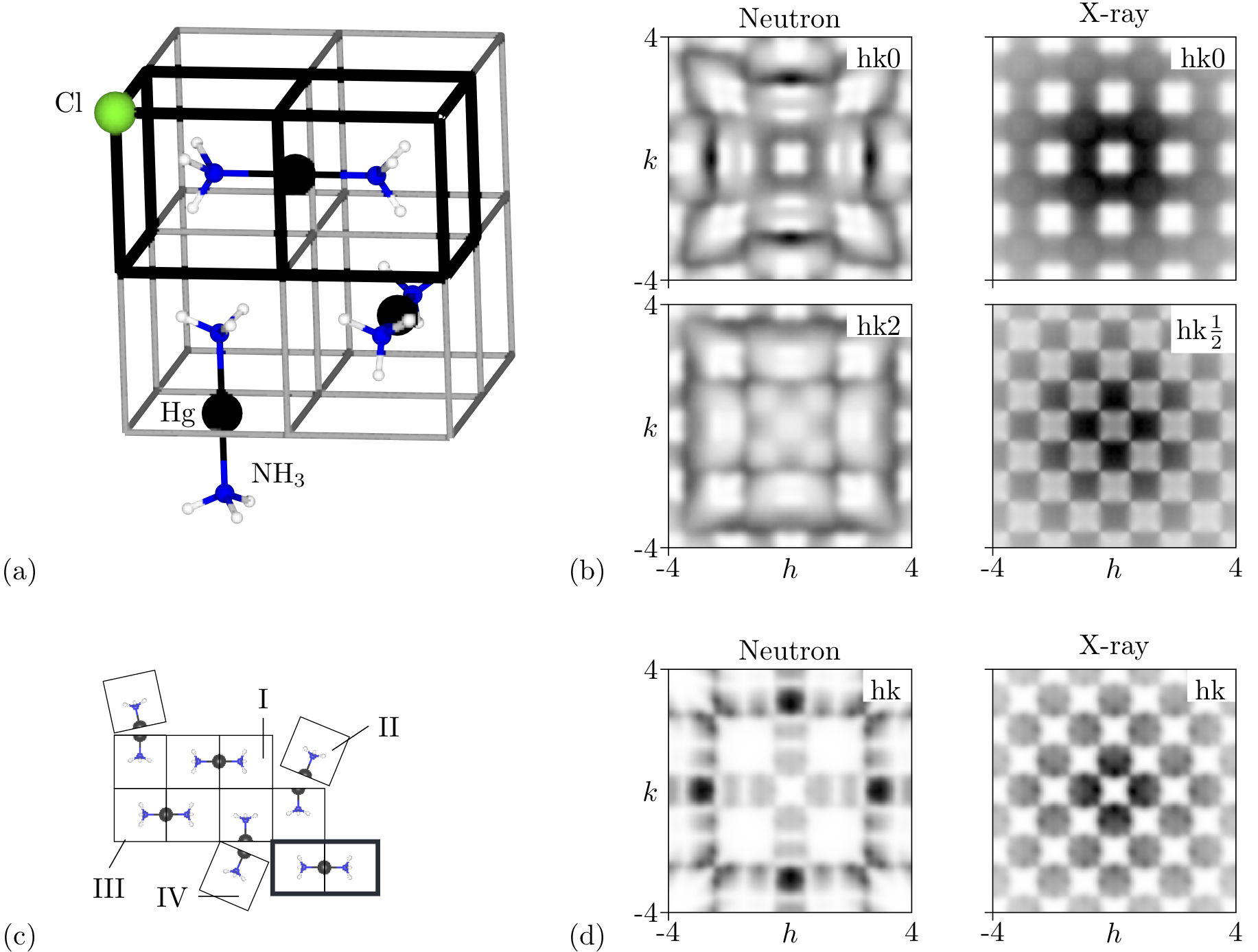}
		\caption{(a) Structural model of Hg(NH$_{3}$)$_{2}$Cl$_{2}$ illustrating a possible distribution of the Hg$^{2+}$ ions to form [H$_3$N--Hg--NH$_3$]$^{2+}$ molecules. The Cl$^-$ ions occupy the cube vertices (only one such ion is shown) (b) Slices of the diffuse neutron and X-ray scattering as calculated from our simulated model structures of Hg(NH$_{3}$)$_{2}$Cl$_{2}$. Simulation details are provided in the supplementary information. (c) Simplified two-dimensional structural analogue of Hg(NH$_{3}$)$_{2}$Cl$_{2}$, obtained by projecting onto two spatial dimensions. The four $\langle10\rangle$-oriented tiles of [Hg$_{1/2}$--NH$_3$]$^{+}$ are labeled and illustrate the matching rules of the system. (d) Two-dimensional diffuse neutron and X-ray scattering of the two-dimensional toy model described in (c).}
        \label{fig2}
	\end{figure*}

For the purposes of this study, we use as our model system the compound Hg(NH$_{3}$)$_{2}$Cl$_{2}$ \cite{Lipscomb_1953}, chosen because it exhibits strongly correlated disorder that arises from particularly simple local interactions \cite{Parsonage_1978,Simonov_2020b} [Fig.~\ref{fig2}(a)]. In the physical material, the Cl$^-$ ions are positioned on a simple cubic lattice. At the centre of each Cl$_8$ cube lies an ammonia molecule, oriented such that its electric dipole points along one of the six $\langle 100\rangle$ directions. Pairs of NH$_3$ molecules in neighbouring cells are connected by Hg$^{2+}$ ions to form [H$_3$N--Hg--NH$_3$]$^{2+}$ ions that each span a $2\times1\times1$ `brick' of the cubic Cl$^-$ lattice. Hence the orientations of NH$_3$ molecules within a single brick are strongly coupled (they point towards one another if connected by a common Hg$^{2+}$ ion); however, this local rule is not sufficiently strong as to drive long-range orientational order. The system is instead an example of a `procrystalline' material (label `C$_1$' in the notation of Ref.~\citenum{Overy_2016}), with $Pm\bar3m$ average symmetry.

From a chemical perspective, the structural complexity of Hg(NH$_{3}$)$_{2}$Cl$_{2}$ arises from strongly correlated Hg$^{2+}$ occupancy disorder on the $3c$ Wyckoff position, which couples to orientational disorder on the NH$_3$ ($1b$) position. Because we develop our mean-field formalism in the context of (pure) orientational disorder, we divide the [H$_3$N--Hg--NH$_3$]$^{2+}$ molecules in two and recast the underlying degrees of freedom in terms of orientations of fictitious (but useful) [Hg$_{1/2}$--NH$_3$]$^{+}$ half-molecules. This approach of establishing one-to-one mappings between orientational and compositional disorder problems is well established in the field \cite{Parsonage_1978,Simonov_2020b}.

\subsection{Single-crystal diffuse scattering}

We generated single-crystal X-ray and neutron diffuse scattering patterns for Hg(NH$_{3}$)$_{2}$Cl$_{2}$ by direct calculation from ensembles of explicit atomistic configurations. The configurations themselves were generated using a `loop-move' MC algorithm \cite{Melko_2001,Evertz_2003}, which enabled efficient sampling of the manifold of configurations strictly obeying the [Hg$_{1/2}$--NH$_3$]$^{+}$ half-molecule matching rules. A total of 50 ground-state configurations were generated in this way, each corresponding to a $40\times40\times40$ supercell of the $Pm\bar3m$ unit-cell. Local bond lengths were taken from the related structure of HgNH$_2$Cl \cite{Lipscomb_1953}, and we included rotations of the NH$_{3}$ molecules around the N--Hg bond axis as identified from infrared spectroscopy measurements \cite{Ebisuzaki_1982}. To speed up calculation, we exploited the fast Fourier calculation algorithm developed in Ref.~\citenum{Paddison_2019}. We further improved the smoothness of the calculated $I(\boldsymbol q)$ functions by applying $m\bar3m$ Laue symmetry. Key slices of the diffuse scattering patterns are shown in Fig.~\ref{fig2}(b); full details of our calculations are given in the supplementary information.

\subsection{Two-dimensional toy model}

It will suit our purposes to establish proof-of-principle of the mean-field approach using a simplified model of Hg(NH$_{3}$)$_{2}$Cl$_{2}$, obtained by projecting onto two spatial dimensions. Because the Cl$^-$ ions are ordered, we omit them altogether from this model. The six $\langle100\rangle$ [Hg$_{1/2}$--NH$_3$]$^{+}$ orientations of the parent 3D model are replaced by four $\langle10\rangle$-oriented tiles in two dimensions [Fig.~\ref{fig2}(c)]; equivalent matching rules apply. Formally, the set of fully-matched tilings now corresponds to the `S$_1$' procrystalline model \cite{Overy_2016} and maps to the gound-state of the `square dimer' model \cite{Kasteleyn_1961}.
	
One particular advantage of a 2D model is that the corresponding diffuse scattering pattern is more straightforwardly presented. We show in Fig.~\ref{fig2}(d) the X-ray and neutron $I(\boldsymbol q)$ functions calculated for the reciprocal-space region $-4\leq h,k \leq4$. This calculation was again based on loop-move-derived MC configurations (50 multiples of $100\times100$ supercells), but was now carried out using the DISCUS program \cite{Neder_2008}. The simulation procedures are described in more detail in the supplementary information.
	
\subsection{Articulating the challenge to be addressed}
     
Our key goal is to establish whether, having measured the diffuse scattering patterns of the type shown in Fig.~\ref{fig2}(b) or (d), we can use a mean-field-based approach to extract the physical interactions responsible for strongly-correlated disorder in the relevant 2D or 3D systems --- in other words, to recover the underlying matching rules that ultimately provide a succinct interaction-space description of the complex real-space order. This contrasts both atomistic and correlation function approaches, against which we will come to compare our results. We start in 2D, progress to 3D, and conclude our results by assessing resilience to data loss.

\section{Results}\label{results}

\subsection{Proof-of-concept: 2D toy model}

\subsubsection{Viability of mean-field approach}

Before attempting a mean-field-based refinement of diffuse scattering data, we must first establish that the $I(\boldsymbol q)$ function calculated using Eq.~\eqref{eq:ID2} for a sensible interaction model actually provides an accurate representation of the explicit result for our 2D toy model [Fig.~\ref{fig2}(d)]. Consequently we assemble a set of interaction matrices based on penalising neighbour-tile pairings that violate the matching rules described above. Consider, by way of an example, the forbidden neighbours in the $(1,0)$ direction for the tile labelled `III' in Fig.~\ref{fig2}(c): each of tiles `II', `III' and `IV' would leave an un-matched half-Hg and hence we assign a penalty $j>0$ to these pairings. By contrast, it is only a `I' tile that is forbidden from neighbouring another `I' tile in this same orientation. Enumerating all possibilities, we arrive at the interaction matrix
	\begin{equation}
    \underline{\underline{J}}_{(1,0)} = \begin{pmatrix} j & j & 0 & j \\
	0 & 0 &j & 0 \\
	0 & 0 &j & 0 \\
	0 & 0 &j & 0  \end{pmatrix}.
	\label{eq:HamiltonDirect}
	\end{equation}
Equivalent matrices for the $(-1,0)$, $(0,1)$ and $(0,-1)$ directions are generated straightforwardly by symmetry.

\begin{figure}
        \includegraphics[width=\columnwidth]{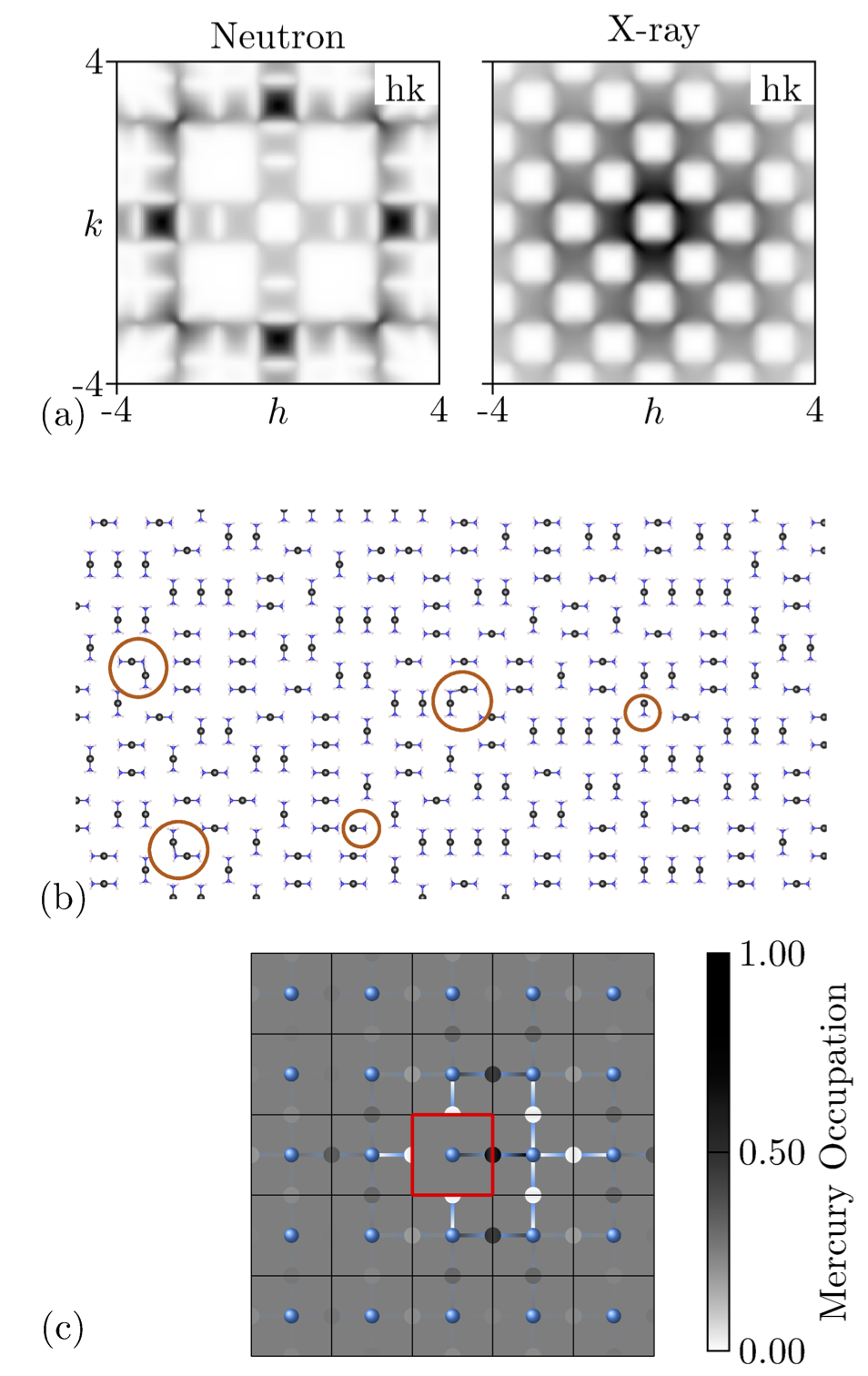}
		\caption{(a) Two-dimensional neutron and X-ray diffuse scattering calculated from the mean field refinement. (b) Section of a sample configuration generated by a direct MC simulation driven by the pair-interaction Hamiltonian of Eq.~\eqref{eq:HamiltonDirect} with $\beta j = 1.617$. Instances where the tiles are not matched correctly are circled. (c) Correlation function of the configuration shown in (b) averaged for symmetry. A tile orientation III is chosen as a reference, indicated by the red box. The probability of the different orientations of the neighbouring tiles are indicated by the occupation-probability of the Hg$_{1/2}$ according to the black and white scale shown on the right-hand side. The ammonia groups are indicated in blue for reference. The nearest-neighbour matching rules are fulfilled to a great extent, as can be seen by the white circles, representing forbidden orientations, that surround the red box.}
        \label{fig3}
	\end{figure}

With access to the $\underline{\underline{J}}$, and using the starting value $\beta j=0$, Eq.~\eqref{eq:ID2} can be used to calculate $I(\boldsymbol q)$ and a goodness-of-fit to the reference data --- both X-ray and neutron --- calculated. The value of $\beta j$ was subsequently refined using least-squares minimisation; note that the optimal scale-factor $\gamma$ can be derived analytically at each step of the minimisation. All derivatives were calculated numerically and $\left| \frac{\Delta j}{j}\right| < 10^{-3}$ was used as the convergence criterion. The refinement was repeated using all 50 model data sets, allowing us to obtain estimates of the parameter uncertainties; full details of our refinements are given as supporting information. Visual inspection of the calculated diffuse scattering patterns makes clear that the mean-field approach is indeed capable of capturing its form [Fig.~\ref{fig3}(a)]. The corresponding $R$-values were 13.4(7) and 14.6(9)\% for, respectively, neutron and X-ray scattering patterns. Both data sets gave very similar values of $\beta j$, each of which satisfies the mean-field approximation criterion: 1.750(4) (neutron) and 1.617(9) (X-ray). We consider it remarkable that such high quality fits can be achieved using just one parameter.
	
Of course a key advantage of this `interaction-space' solution is that --- despite its terseness --- it can nonetheless be used to generate a real-space realisation of the model of arbitrary physical size. We used MC simulations, driven by the interaction matrices exemplified by Eq.~\eqref{eq:HamiltonDirect} and the mean-field value of $\beta j = 1.617$, to obtain a series of atomistic configurations of the 2D system. A region of one such configuration is illustrated in Fig.~\ref{fig3}(b); what is clear is that the matching rules are indeed well obeyed (the exact extent to which naturally depends on the MC temperature). These configurations can also be used to calculate the orientational correlation function, shown schematically in Fig.~\ref{fig3}(c). Note that non-vanishing correlations are observed for longer-range neighbour-pairs beyond those included in the interaction Hamiltonian. We will return to this point in due course.

\subsubsection{Model-agnostic mean-field refinement}

In assembling the interaction matrices above [\emph{e.g.}\ as in Eq.~\eqref{eq:HamiltonDirect}] we have exploited our \emph{a priori} understanding of the interactions between neighbouring tiles; consequently, our next step is to ascertain whether the form of the interaction matrix might itself be determined by refinement against diffuse scattering data. It can be shown that there are exactly seven symmetry-inequivalent terms in the nearest-neighbour tile-pair interaction matrices; the universal form for the $(1,0)$ direction, by way of example, is
	\begin{equation}
    \underline{\underline{J}}_{(1,0)} = \begin{pmatrix} j_{11} & j_{12} & j_{13} & j_{12} \\
	j_{21} & j_{22} &j_{12} & j_{24} \\
	j_{31} & j_{21} &j_{11} & j_{21} \\
	j_{21} & j_{24} &j_{12} & j_{22}  \end{pmatrix}.
	\label{eq:generalhamil}
	\end{equation}
As before, equivalent matrices for the $(-1,0)$, $(0,1)$, and $(0,-1)$ directions are obtained by symmetry; no additional $j_{xy}$ terms are required beyond those in Eq.~\eqref{eq:generalhamil}. 

\begin{figure}
        \includegraphics[width=0.95\columnwidth]{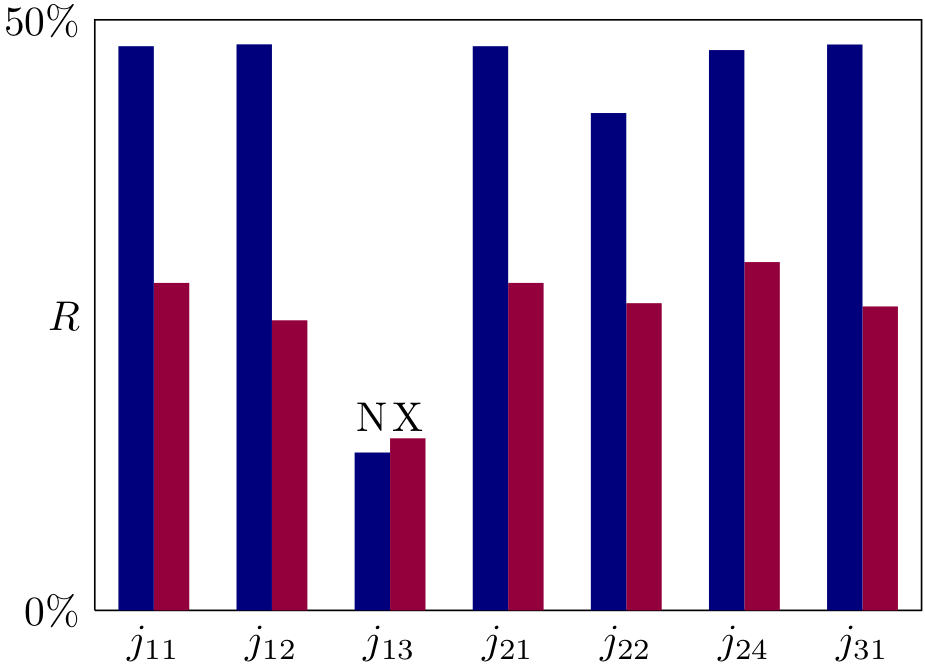}
		\caption{Variation in $R$-values for the model-agnostic mean-field refinement with the different $j_{xy}$ of Eq.~\eqref{eq:generalhamil}. Neutron scattering (N) in blue and X-ray scattering (X) in magenta.}
        \label{fig4}
	\end{figure}

We used our model data sets to carry out a series of mean-field refinements in which we fixed all but one of the $j_{xy}$ parameters to be zero. The quality of fit to the neutron and X-ray diffuse scattering patterns as a function of the single $j_{xy}$ parameter allowed to refine is represented graphically in Fig.~\ref{fig4}. In six of the seven cases, the fits obtained are poor. But in the case that $j_{13}$ is allowed to refine then we obtain fits of exactly the same quality as in our test-case above --- the calculated diffuse scattering patterns in the two instances are indistinguishable (see SI). It can be shown analytically that Eqs.~\eqref{eq:HamiltonDirect} and \eqref{eq:generalhamil} lead to the same eigenvalue problem when $j_{13}=-2j$ and $j_{xy}=0$ otherwise. We indeed find the refined values of $j_{13}$ to be equal to $-2j$. In retrospect, this equivalence is straightforward to understand: the underlying physics responsible for complexity in this system can be couched either in terms of penalising forbidden tile-pairs ($j=j_{11}=j_{12}>0$) or, equivalently, in terms of rewarding matching tile-pairs ($j_{13}<0$) in favour of all other possibilities.

In principle, one might develop the refinement further by allowing one or more $j_{xy}$ terms --- in addition to $j_{13}$ --- to refine, or by including interactions beyond nearest-neighbours. In this case, of course, doing so does not improve the fit-to-data further than the optimal solution we have already found. As flagged in Ref.~\citenum{Paddison_2013}, this is the point at which one might claim to have `solved' the problem of identifying the key interactions to which the diffuse scattering data are sensitive.

\subsubsection{Comparison with established refinement approaches}

By this point, we have shown that the single-crystal diffuse scattering patterns of Fig.~\ref{fig2}(d) can be fitted using a mean-field approach with a single interaction parameter, and that the corresponding interaction model can in turn be used to generate atomistic representations of the underlying disordered structure and also the corresponding correlation functions. We now compare these results to those obtained using conventional refinement strategies that aim to fit the diffuse scattering data directly in terms of atomistic configurations, on the one hand, and correlation functions, on the other hand.

We used a custom RMC code to refine atomistic configurations against the 2D neutron and X-ray diffuse scattering patterns of Fig.~\ref{fig2}(d). Each RMC configuration represented a $20\times20$ supercell (\emph{i.e.} 400 orientational parameters); we used 10 independent runs for each of the 50 model data sets, and exploited the fast Fourier transform algorithm of Ref.~\citenum{Paddison_2019} in our diffuse scattering calculations. While the quality of fit-to-data is much better than for the mean-field approach ($R=6.6(1)$\% (neutron) and $R=5.2(1)$\% (X-ray)) --- perhaps unsurprising given the 400-fold increase in number of parameters --- the fraction of correctly-matched tiles is only 75.8(1)\% for the neutron data, and as low as 37.9(1)\% for the  X-ray data. This difference in sensitivity is to be expected, since in the neutron case the H atoms and the off-centred N-atoms have a significant contribution to the molecular form factor, but in the X-ray  case the variations of the molecular form factor in reciprocal space are much weaker because the scattering is dominated by Hg. In either case, the underlying physics responsible for the diffuse scattering is more difficult to extract from these refinements --- and arguably impossible from the X-ray scattering alone --- than for the mean-field approach developed above.

\begin{figure}
        \includegraphics[width=0.95\columnwidth]{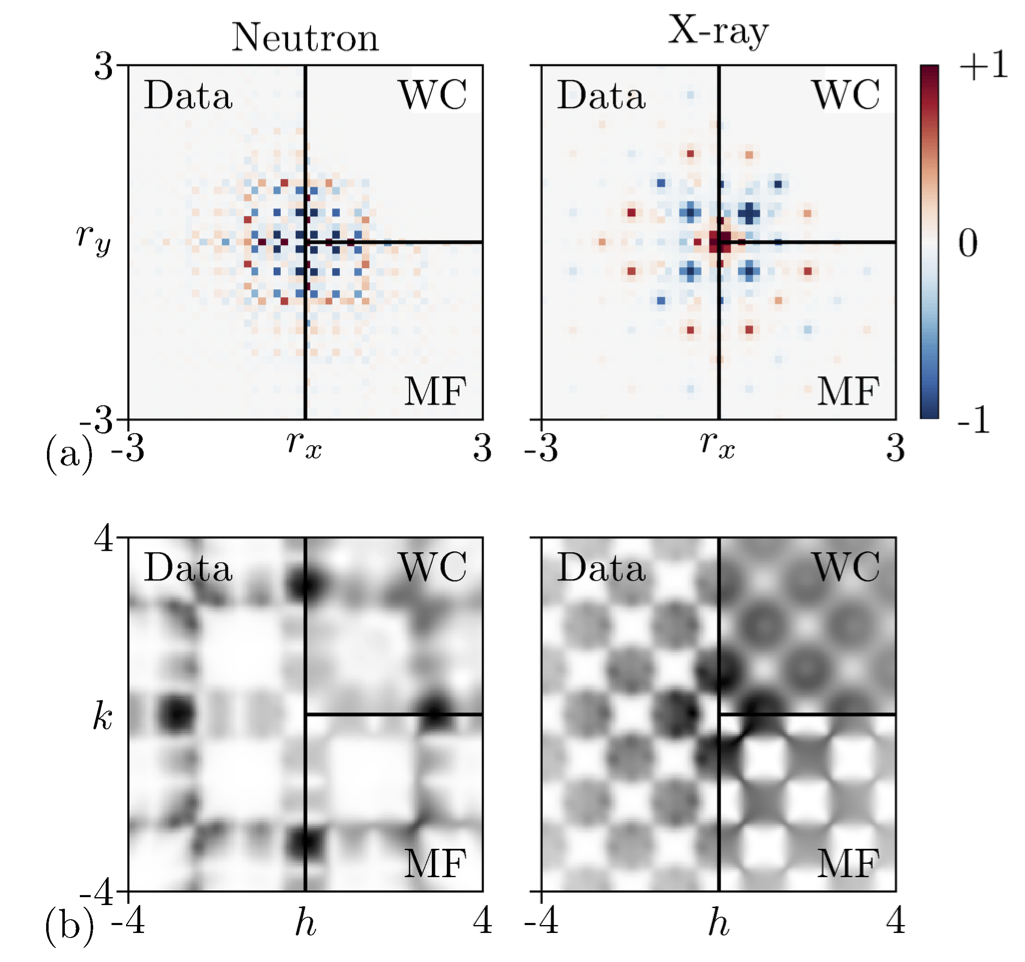}
		\caption{(a) Comparison of the 2D-$\Delta$PDFs calculated from the diffuse scattering of the model data (left), the Warren--Cowley refinement (top right), and mean-field refinement (bottom right). Positive correlations are indicated in red, and negative correlations in blue. (b) Comparison of diffuse scattering used to calculate the 2D-$\Delta$PDFs in (a).}
        \label{fig5}
	\end{figure}

The pairwise correlations from which the diffuse scattering arises can be accessed through the inverse Fourier transform of the diffuse scattering intensities; the corresponding function, known as the 2D-$\Delta$PDF \cite{Simonov_2014}, is weighted by the X-ray/neutron atomic scattering factors and is shown in Fig.~\ref{fig5}(a) for our toy model. Some interpretation of these functions is possible directly; for example, the negative peak at ${\boldsymbol r}\in\langle\frac{1}{2},\frac{1}{2}\rangle$ in the X-ray 2D-$\Delta$PDF implies that Hg--Hg contacts at these vectors are forbidden. Quantitative refinement of the 2D-$\Delta$PDF is possible using the YELL code \cite{Simonov_2014b}. We used a suitably customised version \cite{Schmidt_2017} to refine Warren--Cowley correlation parameters for nearest-neighbours; by exploiting various symmetry relations there are just three independent parameters to be determined (full details are given in the supplementary information). These parameters effectively define the probability of different tile-pairs, and our results correspond to a better-than-RMC --- but not yet perfect --- observation of the original matching rules: these rules being obeyed 87.6(5)\% and 76.8(57)\% of the time for neutron and X-ray data, respectively. In reciprocal space, the corresponding diffuse scattering patterns provide a reasonable match to data ($R=19.9(2)$\% (neutron) and $R=15.4(9)$\% (X-ray)), but are much broadened because they are generated using correlations only at small-$\boldsymbol r$ values [Fig.~\ref{fig5}(b)]. A better fit-to-data would require additional Warren--Cowley terms to be included in the refinement.
   
An important point is made by comparing the 2D-$\Delta$PDFs obtained using our mean-field approach, on the one hand, and those represented by this YELL (Warren--Cowley) refinement, on the other hand. Recall that the former arises from refining nearest-neighbour \emph{interactions} and the latter from nearest-neighbour \emph{correlations}. The mean-field 2D-$\Delta$PDF is structured beyond the nearest-neighbour positions, because short-range interactions can nonetheless affect longer-range structure (a famous example being the order-by-disorder transition in hard-sphere fluids \cite{Alder_1957,Hoover_1968}). So despite including fewer refineable parameters, the mean-field approach actually gives rise to a more detailed structural model. This is the nub of our argument in favour of an `interaction-space' refinement strategy, as exemplified in the mean-field approach we present here.
    
\subsection{3D}

We turn now to the arguably more physical problem of correlated disorder in the three-dimensional procrystal Hg(NH$_3$)$_2$Cl$_2$. Our approach follows closely that described above for the 2D toy model, and we find ourselves able to draw essentially the same conclusions. We summarise below the key results of mean-field, RMC, and Warren--Cowley refinements against both X-ray and neutron simulated diffuse scattering data. In each case the square tiles of the 2D model are replaced by cubic `blocks' that correspond to the six possible orientations of [Hg$_{1/2}$--NH$_3$]$^+$ half-molecules to be arranged within the simple-cubic Cl$^-$ sublattice.

\begin{figure}[b]
        \includegraphics[width=0.95\columnwidth]{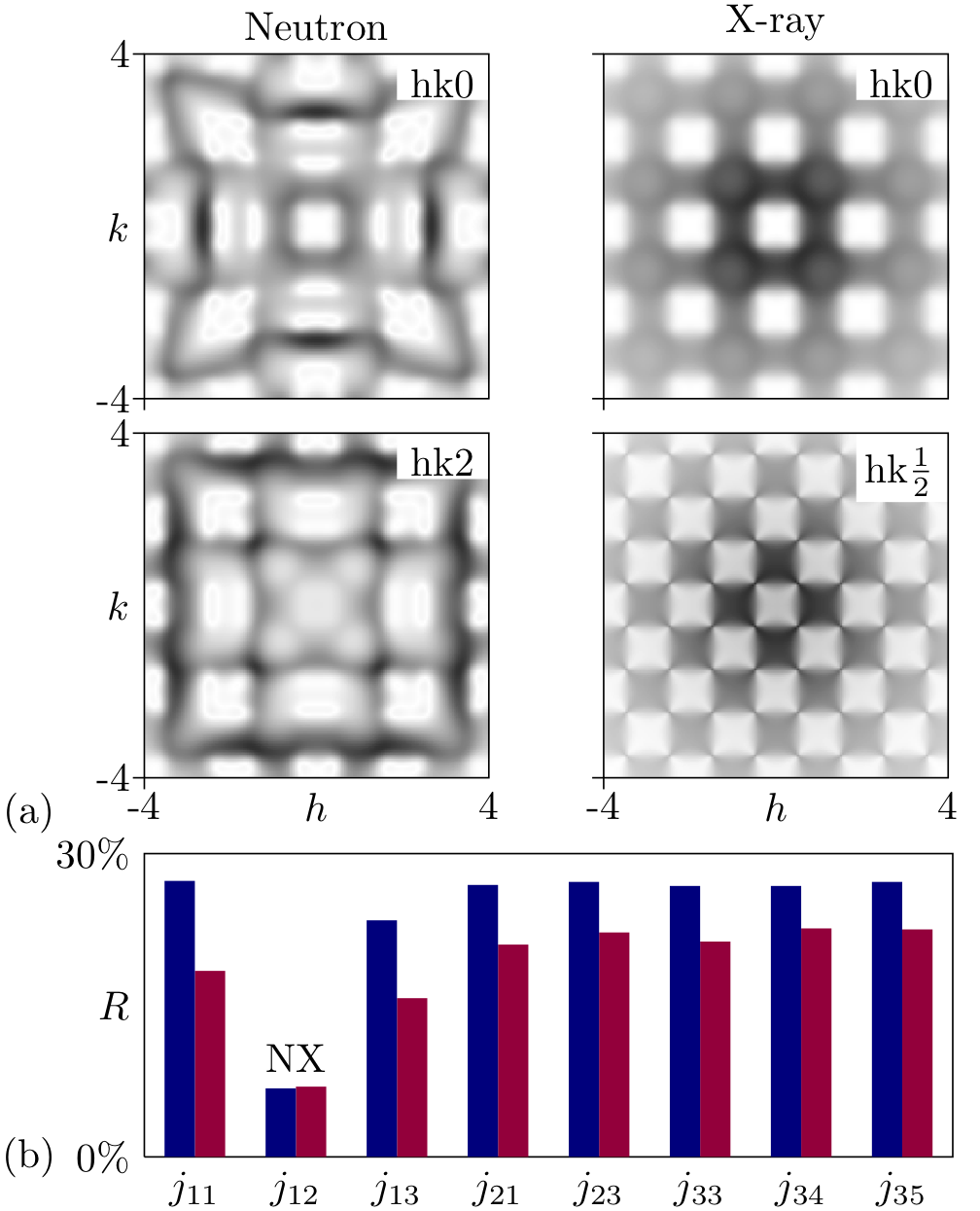}
        
		\caption{(a) Slices of the three-dimensional diffuse neutron and X-ray scattering 
		calculated from our mean-field refinement. (b) Variation in $R$-values for the model-agnostic mean field refinement of the three dimensional data with the different $j_{xy}$ (see supplementary information for further details on the nomenclature and the complete pair-interaction Hamiltonian). Neutron scattering (N) in blue and X-ray scattering (X) in purple.}
        \label{fig6}
	\end{figure}

The mean-field equation~\eqref{eq:ID2} gives an excellent representation of both neutron and X-ray diffuse scattering patterns for nearest-neighbour interaction matrices that assign a single common energy penalty $j$ to mismatched neighbour-blocks. The refined values of $\beta j=2.518(3)$ (neutron) and 2.959(10) (X-ray) satisfy the mean-field approximation criteria, and the remarkable goodness of fit values $R=6.7(2),4.2(3)\%$ are better even than for the 2D model. Morever, the data themselves can be again used to determine the form of the interaction matrices without presuming their form. There are now eight independent entries of the interaction matrices, and only one ($j_{12}$) is needed to obtain high quality fits to data [Fig.~\ref{fig6}]. This key parameter encodes the matching rules and can be used to drive MC simulations that obey these rules (see SI for further discussion). So again we conclude that --- with a single refined parameter --- the mean-field approach can extract from either neutron or X-ray diffuse scattering data the microscopic mechanism responsible for complexity in this representative system.
    
 RMC refinements are somewhat less successful, and especially so for the case of X-ray diffuse scattering patterns. Our series of ten independent refinements on ten of our model data sets used $6\times6\times6$ supercells (216 parameters) yet the fraction of correctly-matched tiles was 82.22(3)\% when driven by the neutron diffuse scattering data, and only 21.44(3)\% for the X-ray data. Hence the additional degrees of freedom in RMC \emph{vs} mean-field approaches serve simply to open up a large configurational space of models that are unphysical yet nonetheless capable of reproducing aspects of the experimental diffuse scattering patterns. The full results of these refinements and the corresponding fits-to-data are given as supporting information.

Finally, we also carried out a Warren--Cowley correlation parameter refinement, confining ourselves to nearest-neighbour terms. There are six independent variables involved in these fits; full details are provided in the supporting information. For the same reason as discussed in the context of our 2D model, the calculated diffuse scattering is necessarily broadened with respect to experiment, but the matching rules are much more faithfully observed than in the RMC case: the refined correlation parameters correspond to 96(3)\% and 83(7)\% of correctly-matched tiles for neutron and X-ray data, respectively.

Across these three different refinement strategies, the best fits to data, the most favourable data-to-parameter ratio, and the clearest path from measurement to identifying the underlying physics responsible for complexity are again all given by the mean-field approach.

    \subsection{Stability against missing data}	
    
Since the mean-field formalism allows such a parameter-efficient means of fitting diffuse scattering data, we sought to establish the extent to which the approach might tolerate incomplete data. We have some experience in this regard from earlier studies of magnetic diffuse scattering, where 2D slices of the full 3D magnetic diffuse scattering pattern have been shown sufficient to refine robustly the corresponding spin interaction model \cite{Paddison_2013}. In the case of conventional (non-magnetic) scattering it is usually possible --- if time-consuming --- to measure complete single-crystal diffuse scattering patterns, at least under ambient conditions \cite{Welberry_2016b}. However, the use of sample environments --- \emph{e.g.} diamond-anvil cells for high pressure measurements \cite{Katrusiak_2008}, gas cells \cite{Yufit_2005}, electric-field cells \cite{Gorfman_2013} --- often imposes severe constraints on reciprocal space coverage. Likewise, even if 3D reciprocal space is well covered in a measurement, it can be difficult to determine accurate diffuse scattering intensities close to the Bragg reflections; this is the motivation for so-called `punch and fill' algorithms \cite{Kobas_2005}, which intentionally discard data in these regions of the diffraction pattern. Consequently the ongoing challenge of linking structural complexity to material function may be aided by the development of efficient refinement strategies that are robust to partial data loss.

We used our 2D toy model and its neutron/X-ray diffuse scattering functions to test the implications of data loss in three cases:
\begin{enumerate}
\item{The omission of scattering intensities at and near Bragg reflections, mimicking the `punch' element of the punch-and-fill approach;}
\item{The restriction of data to a 10$^\circ$ wedge, such as might be encountered when using a diamond-anvil cell; and}
\item{The limiting case of a one-dimensional cut, taken along the $(h0)$ reciprocal-space axis.} 
\end{enumerate}
For each scenario we carried out a mean-field refinement employing the interaction matrices related to Eq.~\eqref{eq:HamiltonDirect} and illustrate the success of this refinement by attempting to reconstruct the full 2D diffuse scattering pattern. We then compare this reconstruction against those generated using RMC and Warren--Cowley refinements.

Our results are summarised in Fig.~\ref{Fig:MissingData}, with a full analysis and a discussion of the extension to 3D given in the supplementary information. The key observation is that mean-field refinement is indeed remarkably tolerant to data loss, with even the extreme case of a 1D cut allowing reasonable recovery of the full 2D data set. Remarkably, it is even possible to carry out a model-agnostic refinement and identify the correct Hamiltonian (see SI). We argue that this robustness arises because the symmetry of the interaction Hamiltonian enforces very strong constraints on the form of the $I({\boldsymbol q})$ function. Symmetry also plays a role in stabilising the Warren--Cowley refinements --- although it is now the (weaker) symmetry of the correlation functions that constrains $I({\boldsymbol q})$. By contrast, RMC is notably intolerant to data loss, which is perhaps unsurprising since it is (by design) agnostic about either crystal or interaction symmetry.

	 	 \begin{figure*}
        \includegraphics[width=0.95\textwidth]{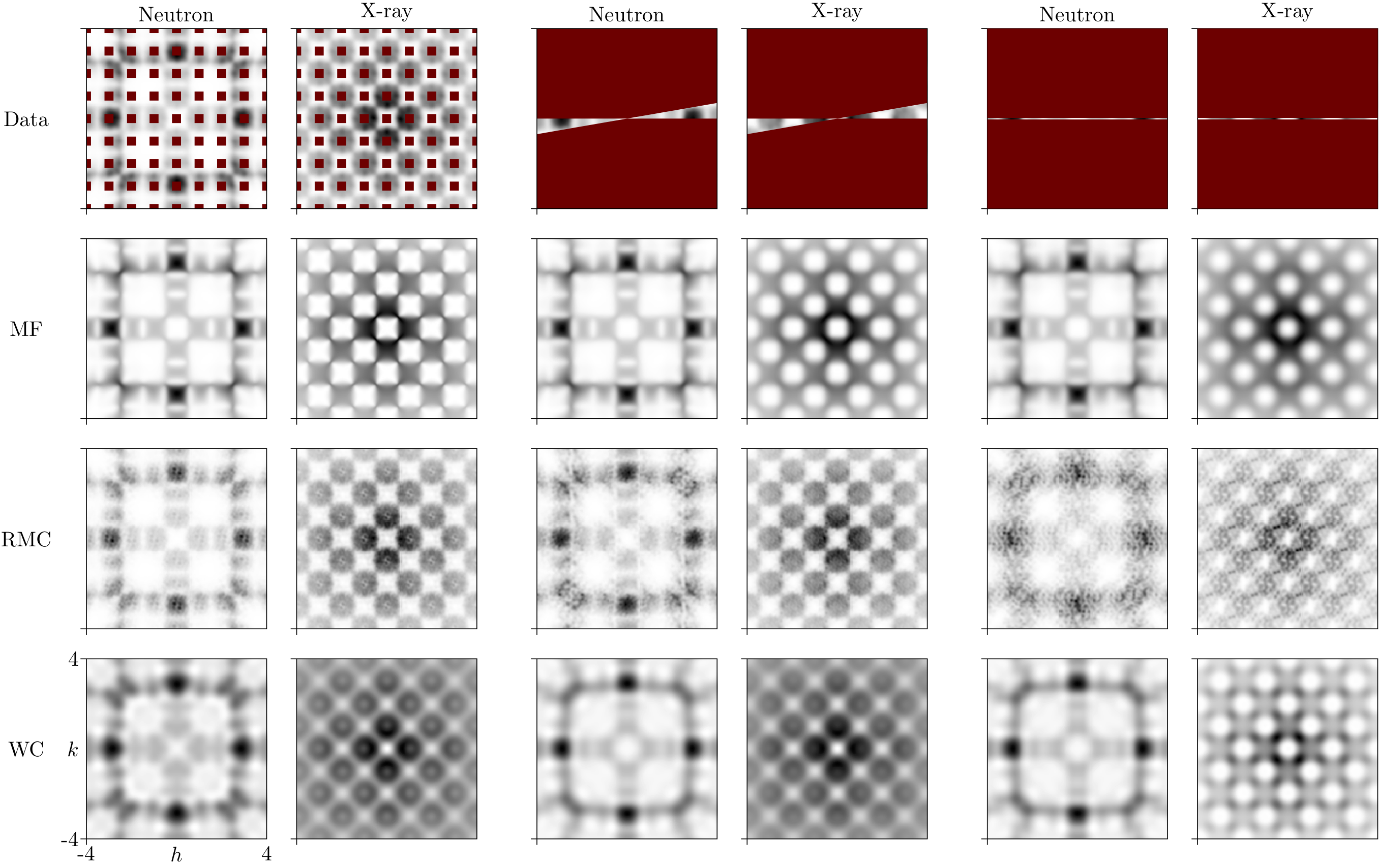}
		\caption{Refinements for missing data input. The top row shows the data input for the three different cases, for both neutron and X-ray diffuse scattering: left, omission of scattering intensities close to the Bragg reflections; middle, restriction of the data to a 10$^{\circ}$ wedge; and right, one-dimensional $h0$ cut. Omitted data are indicated by magenta blocks. Rows two to four show the complete diffuse scattering as recovered from the refinements to the restricted data input for the mean-field refinement (row two), RMC refinement (row three) and Warren--Cowley refinement (bottom row). }
        \label{Fig:MissingData}
	\end{figure*}
	 
	\section{Concluding remarks and outlook}\label{conclusions}
	
For the specific procrystalline system Hg(NH$_{3}$)$_{2}$Cl$_{2}$, and its 2D toy analogue, we have established that a mean-field refinement approach allows unambiguous determination from the diffuse scattering patterns of the microscopic interactions that drive their structural complexity. This link can be established without assuming the form of the interactions, and the process is even robust to (extreme) data incompleteness. We cautiously suggest that the interaction-space approach we have taken here might form the basis for a more general strategy for `solving' the structures of complex and/or disordered crystals \cite{Goodwin_2019}. Certainly the formalism as presented here is not limited to substitutional disorder in molecular systems with homogeneous average occupancies. The molecular form factors can be easily replaced by atomic form factors; and unequal average occupations are straightforwardly accommodated by adapting the matrix $\underline{\underline{M}}$. The formalism is easily extended to allow treatment of crystallographic space groups that contain several disordered sites in the unit cell, and such an extension would allow investigation of nonstoichiometric compounds as described in Refs.~\citenum{Gusev_2006} and \citenum{Withers_2015}.

Looking forward more generally, what challenges might one expect to face? An obvious limitation will be the study of systems poorly described by the mean-field approximation --- \emph{e.g.}\ when the stability criteria are not met, or cases that are far from equilibrium. Likewise, the formalism as described here relies on discrete degrees of freedom, and so is well suited to problems that can be phrased in terms of occupational disorder. The extension to continuous degrees of freedom --- needed to capture particular types of displacive disorder, for example --- is an important challenge that we are hoping to address in the near future. One expects additional difficulties whenever there is nontrivial interplay between various different degrees of freedom, such compositional, displacive, and magnetic. Nevertheless we hope to have demonstrated here that the potential reward for developing a generalised mean-field approach to fitting diffuse scattering data may be great indeed.

\section*{Acknowledgements}
	We thank Joseph Paddison (Oak Ridge) for useful discussions and the ERC (Grant 788144) and the Leverhulme Trust (Grant RPG-2018-268) for funding.

\bibliography{arxiv_2021_mf} 

\end{document}